\documentclass{article}
\usepackage{spconf,amsmath,graphicx}
\usepackage{umoline}
\usepackage{mwe} 
\usepackage{amsmath}
\usepackage{amsthm}
\usepackage{amssymb}
\usepackage{graphicx}
\usepackage{epstopdf}
\usepackage{subcaption}
\usepackage[skip=2pt,font=small]{caption}
\usepackage{adjustbox}
\usepackage{algorithm}
\usepackage{mwe,tikz}
\usepackage[percent]{overpic}
\usepackage{algorithmic,color}
\usepackage{multirow}
\usepackage{appendix}
\usepackage{pgfplots}
\usepackage{xcolor,colortbl}
\usepackage{flushend}
\definecolor{MyCyan}{RGB}{189,242,216}
\definecolor{MyCoral}{RGB}{253,124,110}
\newcolumntype{a}{c}
\newcolumntype{d}{c}
\usepackage{autonum}

\newcommand{\R}{{\mathbb R}}

\newcommand{\Sm}{{\mathcal{S}}}
\newcommand{\Hm}{{\mathcal{H}}}
\newcommand{\Dm}{{\mathcal{D}}}
\newcommand{\Nm}{{\mathcal{N}}}
\newcommand{\al}{\alpha}

\DeclareMathOperator*{\argmin}{arg\,min}

\def\x{{\mathbf x}}

\title{Non-convex Super-resolution of OCT images via sparse representation}
\name{Gabriele Scrivanti$^{\dagger\star}$ \quad Luca Calatroni$^{\star}$ \quad Serena Morigi$^{\dagger}$ \quad Lindsay Nicholson$^{\ddag}$ \quad Alin Achim$^{\ddag\star}$ \thanks{AA acknowledges partial support of the Leverhulme Trust Research Fellowship INFHER, LC and GS the one of the UCA IDEX JEDI grant ``Attractivit\'e du territoire", GS and SM the support of the Dept.Math, UNIBO.}}

 \address{
  $^{\dagger}$ Department of Mathematics, University of Bologna, Italy \\
  $^{\star}$ Laboratoire I3S, CNRS, UCA, INRIA,  Sophia-Antipolis, France \\
  $^{\ddag}$ University of Bristol, UK}

\begin{document}
%
\maketitle
\begin{abstract}
We propose a non-convex variational model for the super-resolution of Optical Coherence Tomography (OCT) images of the murine eye, by enforcing sparsity with respect to suitable dictionaries learnt from high-resolution OCT data.
The statistical characteristics of OCT images motivate the use of $\alpha$-stable distributions for learning dictionaries, by considering the non-Gaussian case, $\alpha =1$. 
The sparsity-promoting cost function relies on a non-convex penalty - Cauchy-based or Minimax Concave Penalty (MCP) - which makes the problem particularly challenging. 
We propose an efficient algorithm for minimizing the function based on the forward-backward splitting strategy  which guarantees at each iteration the existence and uniqueness of the proximal point. 
Comparisons with standard convex $\ell_1$-based reconstructions show the better performance of non-convex  models, especially in view of further OCT image analysis.
\end{abstract}
\begin{keywords}
Optical Coherence Tomography, Super-Resolution, Non-Convex Regularisation, Sparse Representation.
\end{keywords}
\section{Introduction}
\label{sec:intro}

OCT is an \textit{in vivo} non-invasive imaging technique based on low-coherence interferometry that allows to detect ophthalmic structures at micrometer resolution. OCT images show sections of the multiple layers of the retinal tissue as well as the inner eye region (vitreous) so they are particularly suited for the detection of anomalies and deformations in the eyes as well as in the follow-up of ophthalmic diseases in early and later stages such as Multiple Sclerosis, Diabetes, Alzheimer’s disease, Parkinson disease, or Glaucoma \cite{SR_OCT1}. However, the poor spatial resolution and the multiplicative nature of the (speckle) noise observed in OCT images often limits the possibility of an accurate image analysis, which makes the use of both super-resolution  (SR) and denoising/despeckling imaging techniques crucial for the subsequent image analysis \cite{SR_OCT1}, often based on accurate (and often manual) layer segmentation.
In this paper, we propose a non-convex variational framework for the super resolution of real murine OCT images based 
on sparse representations with respect to pre-computed high-resolution OCT dictionaries.  The SR problem is formulated as the inverse problem of retrieving the original high-resolution (HR) image from a given low-resolution (LR) one where we account also for the presence of blur and background noise. We further assume that the entries of the desired HR image are symmetric-$\alpha$-stable random variables and 
thus the OCT image can be well represented by only a few atoms of the given HR dictionary learned by  $\alpha$-stable distributions following \cite{dictionarySparseDT}. In our framework, the super-resolution is performed patch-wise as part of the reconstruction procedure as presented in \cite{Superres_sparsrepr}, differently from the alternative approach presented in \cite{SR_OCT1} for SR of OCT images, where the upscaling is obtained via an interpolation step which is preliminary to the patch extraction.
We enforce sparsity by means of a non-convex regularisation term, and, in particular, we consider the separable Cauchy-based penalty and the MCP, which both depend on a scalar positive  parameter which modulates the non-convex behaviour.   Following \cite{Conv:Guar}, we efficiently solve the non-convex problem by means of a forward-backward splitting algorithm \cite{CombettesWajs2005} where the existence and uniqueness of the proximal point are guaranted by suitable conditions introduced on the model and algorithmic parameters. This is different from the recent Convex-Non-Convex (CNC) approaches \cite{LMS15} where conditions on the overall convexity of the total objective functional are derived, while preserving the non-convex behaviour of the regularisation term. The proposed strategy, working in a pure non-convex regime, guarantees, on the other hand, convergence to a stationary point. 
We show quantitative and qualitative comparisons with reconstructions obtained by means of the $\ell_1$-norm based model and different dictionaries.


\section{SR via sparse representation}
The task of recovering a HR OCT image $X \in \R^{r_h \times c_h}$ from a noisy, and blurred LR input $Y\in \R^{r_l \times c_l}$ can be modeled mathematically as an inverse image reconstruction problem whose ill-posedness, as described in \cite{Superres_sparsrepr}, can be overcome by representing $X$ in a sparse way with respect to a given (over-complete) dictionary. The overall problem can be thus modelled by introducing the following constraints. 

1) \textit{Reconstruction Constraint}: the input image $Y$ is linked to the desired HR image $X$ via the model
    \begin{equation}
    Y = \Nm(\Sm_q(\Hm(X))) + \eta
    \tag{C1}
    \label{YSHX}
    \end{equation}
    where $\Hm: \R^{r_h \times c_h}\to \R^{r_h \times c_h}$ is the blur operator corresponding to the point spread function (PSF) of the optical acquisition system considered, $\Sm_q: \R^{r_h \times c_h} \to \R^{r_h/q \times c_h/q}$ is a downsampling operator defined in terms of a factor $q\in\mathbb{N}$ which maps the blurred image $\Hm(X)\in \R^{r_h \times c_h}$ into a coarser grid by averaging. As any imaging technique that is based on detection of coherent waves, OCT images are subject to the presence of speckle noise \cite{Michailovich2006}, thus $\Nm(\cdot)$ stands for a multiplicative (speckle) noise degradation process, which is not considered in this work, while $\eta\in\R^{r_h/q \times c_h/q}$ denotes an additive component which represents white Gaussian background noise.
    
    2) \textit{Sparsity Constraint}: We assume that every (square) patch $x$ extracted from $X \in \R^{r_h \times c_h}$ can be represented as a sparse linear combination of $n_d$ atoms of a given over-complete dictionary $\Dm \in \R^{n_p \times n_d}$ which has been previously learned in terms of HR training images. Using a vectorised notation for any patch $x\in\R^{n_p}$, this assumption translates in:
    \begin{equation}
        x \approx \Dm a\quad \text{for\ }\quad a \in \R^{n_d}\quad\text{with}\quad\|a\|_0 << n_d.
        \tag{C2}
        \label{eq:sparsity_constraint}
    \end{equation}

To learn the dictionary $\Dm$, we consider the approach presented in \cite{dictionarySparseDT} where general $\al$-Stable distributions are used as prior PDFs. In particular, we set the parameter $\al=1$ so that the underlying distribution is assumed to be Cauchy.

We combine \eqref{YSHX} with \eqref{eq:sparsity_constraint} to compute the sparse coefficient vector $a$ for each given LR square patch $y\in \mathbb{R}^{n_p/q^2}$ extracted from the observed image $Y$, 
thus getting the following optimisation problem
\begin{equation}
a^* \in \argmin_{a\in \mathbb{R}^{ n_d}}	\left\{ \frac{1}{2}\| y\!-\! \Sm_q(\Hm(\Dm a))\|_2^2\! +\! \!\lambda\!\sum_{i=1}^{n_d}\phi_\gamma(a_i) \right\},
\tag{P}
\label{eq:cauchy:patch_problem}
\end{equation}
where $\phi_\gamma$ is a numerically tractable version of the $\ell_0$ pseudo-norm in \eqref{eq:sparsity_constraint}  promoting sparsity and depending on a parameter $\gamma>0$, and $\lambda>0$ is a regularisation parameter.

The solution $a^*$ of the optimisation problem \eqref{eq:cauchy:patch_problem} is the sparse coefficient vector representing the HR patch $x\in\R^{n_p}$ 
under the over-complete dictionary $\Dm$.


\section{Sparse non-convex regularisations}  \label{sec:reg}

We describe in this section two popular  sparsity-promoting, non-convex and parametric penalty functions $\phi_\gamma:\R\to \R^+$ in \eqref{eq:cauchy:patch_problem} which are considered in the proposed SR framework. 
Among the non-convex sparsity-promoting regularizers
characterized by tunable degree of non-convexity,
the MCP is considered one of the most interesting and effective penalties \cite{LMS15}. As a natural alternative  
to the MCP we propose the Cauchy-based penalty, assuming that the desired solution vector $a^*$ is distributed as an $n_d$-dimensional Cauchy distribution with parameter $\gamma$;
it is therefore easy to derive an instance of \eqref{eq:cauchy:patch_problem} following, e.g., \cite{Conv:Guar}.

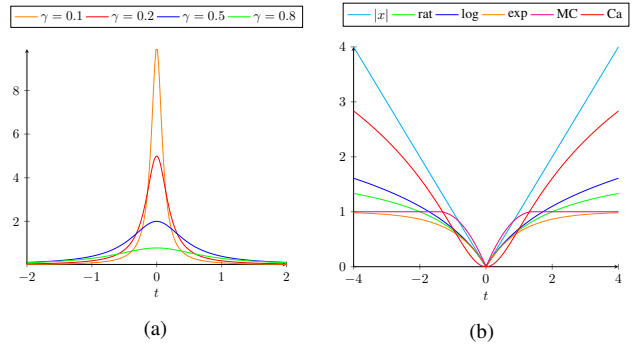
\begin{figure}
\centering
\begin{tabular}{ll}
\begin{subfigure}[h]{0.22\textwidth}
\begin{adjustbox}{width=1\textwidth}
\begin{tikzpicture}
\begin{axis}[
    axis lines = left,
    xlabel = $t$,
    legend style={at={(0.5,1.2)}, anchor=north,legend columns=-1}
]

\addplot [
    domain=-2:2, 
    samples=200, 
    color=orange,
    ]
    {0.1/(x^2 + 0.1^2)};
\addlegendentry{$\gamma = 0.1$}
\addplot [
    domain=-2:2, 
    samples=200, 
    color=red,
]
{0.2/(x^2 + 0.2^2)};
\addlegendentry{$\gamma = 0.2$}
\addplot [
    domain=-2:2, 
    samples=200, 
    color=blue,
    ]
    {0.5/(x^2 + 0.5^2)};
\addlegendentry{$\gamma = 0.5$}
\addplot [
    domain=-2:2, 
    samples=200, 
    color=green,
    ]
    {0.5/(x^2 + 0.8^2)};
\addlegendentry{$\gamma = 0.8$}
\end{axis}
\end{tikzpicture}
\end{adjustbox}
\subcaption{}
\label{fig:cauchy:pdf}
\end{subfigure}
&
\begin{subfigure}[h]{0.22\textwidth}
    \begin{adjustbox}{width=1\textwidth}
    \begin{tikzpicture}
[declare function ={
vararat = 1;
ahalf = vararat / 2;
rat(\x) = abs(\x) / ( 1 + ahalf * abs(\x) );
varalog = 1;
philog(\x) = ( ln( 1 + varalog * abs(\x) ) ) / varalog;
varaexp = 1;
phiexp(\x) = ( 1 - exp( -varaexp * abs(\x) ) ) / varaexp;
alpha=1;
oneover2alpha = 1/(2*alpha);
sqrt2alpha = sqrt(2*alpha);
sqrt2overalpha = sqrt(2/alpha);
mcp(\x) = (-oneover2alpha * abs(\x)^2 + sqrt2alpha*abs(\x))*(abs(\x)<sqrt2overalpha) + 1*(abs(\x)>=sqrt2overalpha) ;
gamma = 1;
cauchy(\x) = ln((gamma^2 + \x^2)/gamma)-ln(gamma);
}
]
\begin{axis}[
    axis lines = left,
    xlabel = $t$,
    legend style={at={(0.5,1.2)}, anchor=north,legend columns=-1}
]
\addplot [
    domain=-4:4, 
    samples=300, 
    color=cyan,
]
{abs(\x)};
\addlegendentry{$|x|$}
\addplot [
    domain=-4:4, 
    samples=300, 
    color=green,
]
{rat(\x)};
\addlegendentry{rat}
\addplot [
    domain=-4:4, 
    samples=300, 
    color=blue,
]
{philog(\x)};
\addlegendentry{log}
\addplot [
    domain=-4:4, 
    samples=300, 
    color=orange,
]
{phiexp(\x)};
\addlegendentry{exp}
\addplot [
    domain=-4:4, 
    samples=300, 
    color=magenta,
]
{mcp(\x)};
\addlegendentry{MC}
\addplot [
    domain=-4:4, 
    samples=300, 
    color=red,
]
{cauchy(\x)};
\addlegendentry{Ca}
\end{axis}
\end{tikzpicture}
    \end{adjustbox}
    \subcaption{}
    \label{fig:penalties}
\end{subfigure}\\
\end{tabular}
\caption{(a) Cauchy PDF for varying $\gamma>0$. (b) Plots of different penalty functions relaxations of the $\ell_0$ pseudo-norm.}
\end{figure}

 The Cauchy distribution belongs to the family of $\al$-stable distributions (with $\alpha=1$) which are heavy-tailed and have been exploited in tomography applications \cite{dictionarySparseDT}. Differently from other probability distributions in the family, in the case of Cauchy distribution there exists a closed-form probability density function whose negative log-likelihood for $t\in\R$ reads:
\begin{equation}
    \phi_\gamma(t) = \phi^{C}_\gamma(t) 
    = \log \left(\frac{\gamma^2 + t^2}{\gamma}\right).
    \label{eq:cauchy:penalty}
\end{equation}
As illustrated in Figure \ref{fig:cauchy:pdf} the parameter $\gamma$ controls the spread of the distribution: the smaller $\gamma$, the narrower and more peaked the shape of the distribution.
We remark that the function $\phi_\gamma$ in \eqref{eq:cauchy:penalty} is non-convex except for a small and limited interval around the origin and increases unbounded at a logarithmic rate. In Figure \ref{fig:penalties} we report the behaviour of the Cauchy penalty when used as a relaxations of the $\ell_0$ pseudo-norm in comparisons with other classical choices as $\ell_1$-norm, $rat$, $log$, $exp$, MCP, see \cite{LMS15} for details. 
Finally, we remark that the only convex function is the $\ell_1$-norm function, which corresponds to  $\phi_\gamma(t)=|t|$.


The Minmax Concave Penalty introduced in \cite{zhang2010} has been successfully applied to several signal and image recovery problems. The function $\phi_\gamma$ in this case reads:
\begin{equation}
\phi_{\gamma}(t) = \phi^{MCP}_\gamma(t)=
    \begin{cases}
    -\frac{1}{2\gamma}t^2 + \sqrt{\frac{2}{\gamma}}t\quad &\text{if}\; |t|<\sqrt{2\gamma},\\
    1 &\text{if}\; |t|\geq \sqrt{2\gamma},
    \end{cases}
    \label{MCP}
\end{equation}
where $\gamma>0$ modulates the concavity of the regulariser.





\section{Forward-backward algorithm}



For every patch, we  solve the non-convex problem \eqref{eq:cauchy:patch_problem} by means of a Forward-Bacwkard (FB) splitting algorithm \cite{CombettesWajs2005}, then we recombine the patches together via sliding averaging. Algorithm \ref{alg:FB-PaBS-OCT} illustrates  the pseudocode where $D$, and $S_q$ denote suitably resized matrices computed by the operators $\Dm$ and $\Sm_q$, respectively. The matrix $H$ represents the PSF of the OCT scan which is estimated  as described in \cite{Michailovich2006}. 

The convergence of the FB algorithm is subordinated to a sufficiently small step size $\mu$ in the forward step [F], and the existence of a unique minimizer of the backward step [B].
Regarding $\mu$, according to \cite{CombettesWajs2005}, the maximal value guaranteeing convergence is $1/L$, where $0<L=\|(S_qHD)^T S_qHD\|$ denotes the Lipschitz constant of the quadratic fidelity term.


In order to have a uniquely defined proximal point at each backward step [B]
we look for a parameter value $\gamma$ such that:
\begin{equation}
    \mathrm{prox}_{\lambda\mu\Phi_\gamma} (x)= \argmin_{u\in \R^{n_d}}{\left\{J(u):=\frac{\|x-u\|_2^2}{2\lambda\mu} + \sum_{i=1}^{n_d}\phi_\gamma(u_i) \right\}}
\end{equation}
is a singleton (i.e. $J(\cdot)$ is strongly convex). By the separability of the penalty $\Phi_\gamma(\cdot):=\sum_{i=1}^{n_d}\phi_\gamma(\cdot)$ we first observe that the proximal point can be computed component-wise on each $i$-th component, $i=1,\ldots, n_d$. Then, we follow the strategy in \cite[Theorem 2]{Conv:Guar} and choose, for both the Cauchy \eqref{eq:cauchy:penalty} and the MCP \eqref{MCP} penalties, a parameter $\gamma$ such that $J$ is strictly 
convex, which guarantees the uniqueness of proximal point.
Balancing the positive second derivatives in the first term (fidelity) against the negative second derivatives in the regularization term of $J$, 
we find that the strict convexity is guaranteed when the following condition holds for $\tau> 1$:
\begin{equation}  \label{eq:conditions}
    \gamma = \tau\bar{\gamma},\quad \bar{\gamma}:=\begin{cases}
    \frac{\sqrt{\lambda\mu}}{2}\quad&\text{if }\quad \phi_{\gamma}=\phi^{C}_\gamma,\\
    \lambda\mu\quad &\text{if }\quad \phi_{\gamma}=\phi_\gamma^{MCP}.
    \end{cases}
\end{equation}
Note that, in both cases, a closed form expression of the proximal points can be easily computed by simply looking for solutions of the corresponding optimality conditions. In the case of Cauchy regularisation the solution can be computed using Cardano's method for solving the cubic optimality condition \cite{Wan2011}, while for MCP a closed formula is proposed in \cite{zhang2010}.

The extension of this algorithmic strategy to the design of converging FB algorithms for general non-convex parametric regularisers $\phi_\gamma$ is left for future research. We remark that \eqref{eq:conditions} does not affect the overall non-convexity of the original problem \eqref{eq:cauchy:patch_problem}. Hence, only convergence to a stationary point can be guaranteed.




For each given LR patch $y$, once the sparse vector $a$ is obtained, the reconstructed HR patch $x$ is computed by enforcing the condition \eqref{eq:sparsity_constraint}. Note that by definition, the HR patch computations are independent between each other and as such they can be parallelised. As a final step, the reconstructed SR image is generated by aggregating all the HR patches together and locally averaging according to the number of overlapping patches at each location. This was shown in \cite{SR_OCT1} to produce reconstruction results with less artefacts.

\begin{center}
\begin{minipage}{0.48\textwidth}
\begin{algorithm}[H]
 \textbf{Input:} $Y$, $D$, $S_q$, $\mu\in (0, 1/ L]$, $\lambda>0$, $\gamma\in (\bar{\gamma},+\infty)$\;
 \textbf{Output:} $X^*$  \textit{\% Super-resolved OCT image}\\
 Estimate PSF as in \cite{Michailovich2006} and get $H$\;
 Extract overlapping patches of size $\frac{\sqrt{n_p} }{q}\times\frac{ \sqrt{n_p}}{q}$\;
 \textbf{For} each patch $y$:\\
 \begin{tabular}{llr}
  & $a_0 = 0$ & \;\\
  &\textbf{For} $k=0,1,\ldots$ \textbf{do:} & \\
  & \quad $b_{k+1} = a_k -\mu D^TH^TS_q^T(y-S_qHDa_k)$ &[F]    \\
  &\quad $a_{k+1} = \text{prox}_{\lambda\mu\Phi_\gamma} (b_{k+1})$ &[B]\\
 &\textbf{until convergence} &\\
 & $x = Da$\; \% \textit{Generate the HR patch} 
 \end{tabular}\\
 \textbf{end}\\
 $X^*$ is obtained by sliding average of the HR patches
 \caption{FB splitting for patch-based sparse OCT SR}
 \label{alg:FB-PaBS-OCT}
\end{algorithm}
\end{minipage}
\end{center}

\section{Numerical Results}
In this section we discuss the performance of the patch-based sparse OCT SR Algorithm \ref{alg:FB-PaBS-OCT} when applied to the case of Cauchy, MCP and $\ell_1$ regularisation and for different choices of the dictionary $D$. The quantitative assessment of the SR performance for all tests is based on the comparison between the obtained SR image $ X^{*} \in \R^{r_h \times c_h}$ and a reference HR version $ X^0 \in \R^{r_h \times c_h}$ of the input image 
in terms of 
Peak Signal to Noise Ratio (PSNR) and
Structural Similarity Index (SSIM). 
We remark that these comparisons are useful to assess only the SR capabilities of the different approaches considered. Due to the lack of reference ground-truth images, in practical situations, however, a qualitative visual evaluation performed by a practitioner is still required to assess the computed results in view of further analysis (typically, layer segmentation).

\begin{figure}
\includegraphics[width=0.45\textwidth]{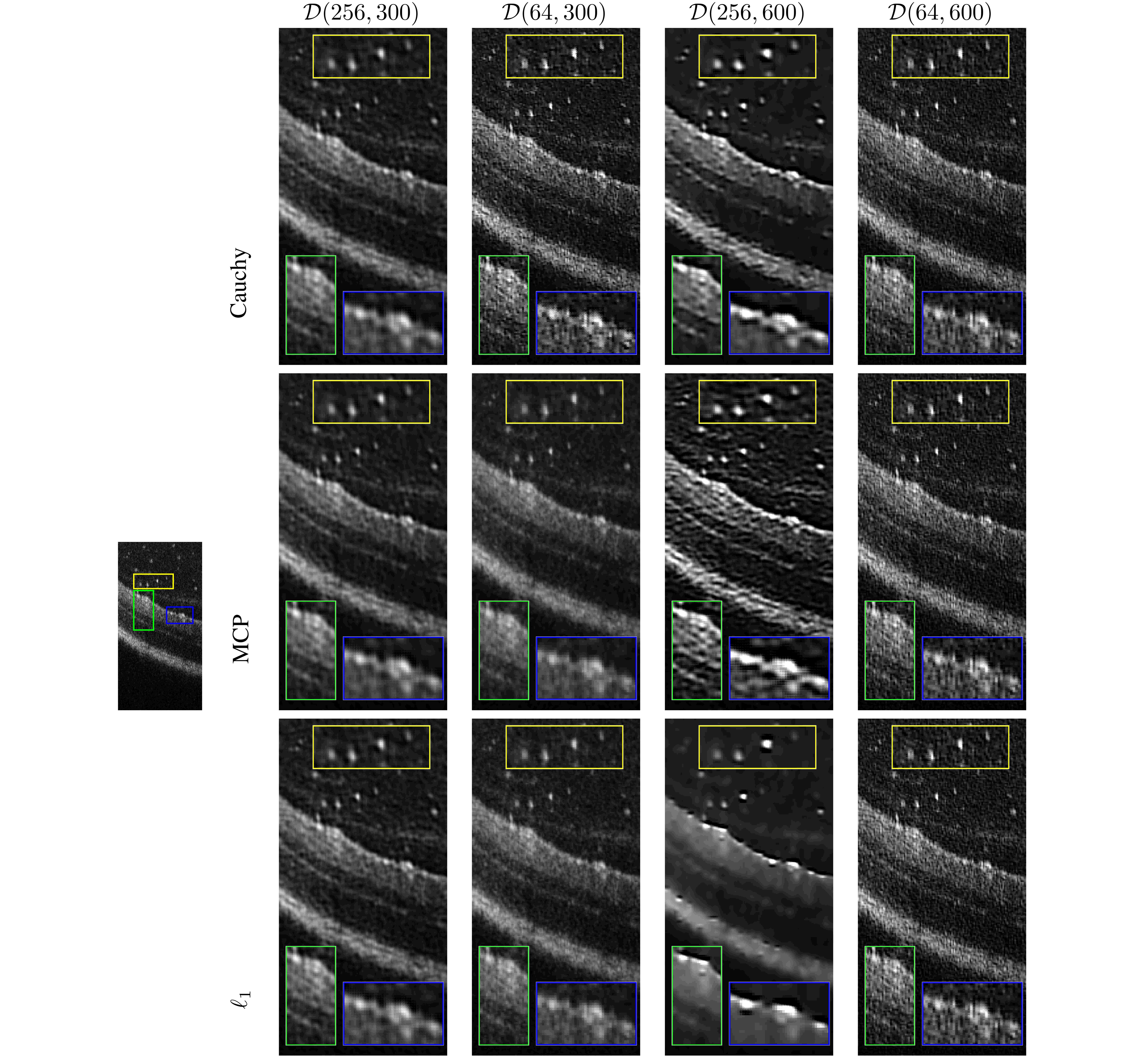}
    \caption{OCT1 Super-Resolution reconstructions ($q=4$).}
    \label{fig:TX12:rec}
\end{figure}

\begin{figure}
\centering
\resizebox{.4\textwidth}{!}{
 
 \begin{tabular}{ccccc}
     \includegraphics[width=1in]{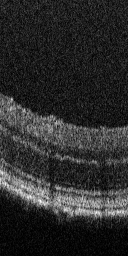}
    & \includegraphics[width=1in]{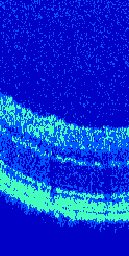}
     &\includegraphics[width=1in]{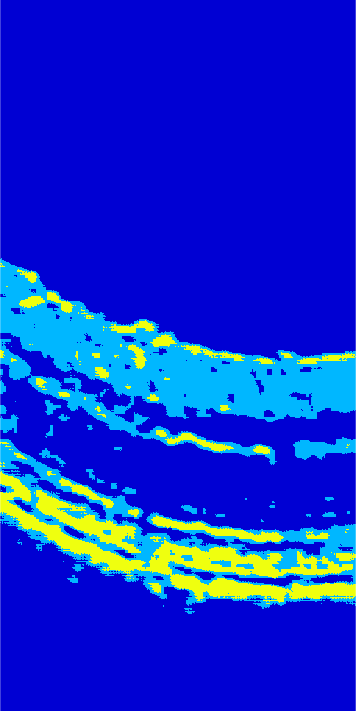}
     &\includegraphics[width=1in]{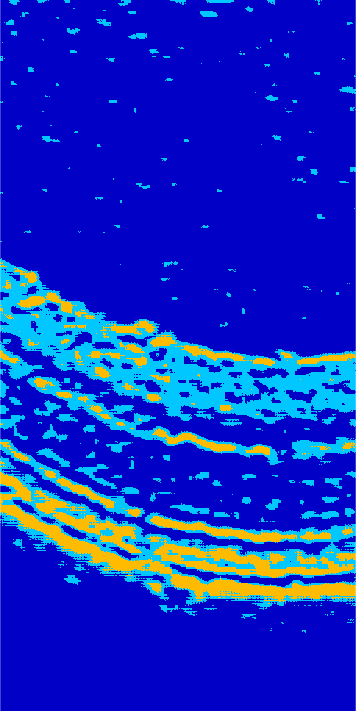}
     &\includegraphics[width=1in]{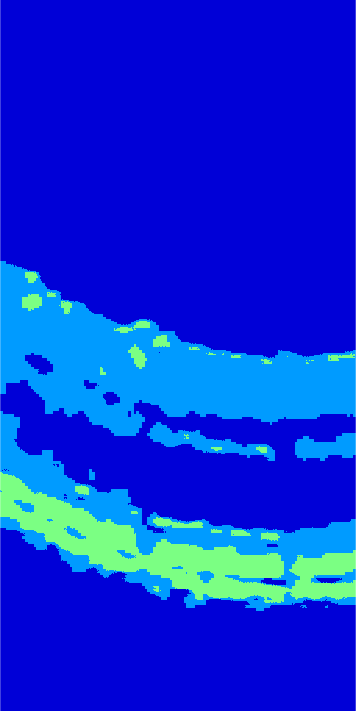}\\
     OCT2 & LR & Cauchy&MCP&$\ell_1$\\
\end{tabular}
}
\caption{OCT2 segmentation with $k$-means before (LR) and after SR (Cauchy, MCP, $\ell_1$), $k=3$, for different regularisations.}
\label{fig:test1:segm}
\end{figure}

\textbf{Dictionaries and parameters.} We considered 4 dictionaries denoted by $\Dm(n_p,n_d)$ which vary according to the number of atoms $n_d \in \{300, 600\}$ and the dimension of each atom $n_p \in\{64,256\}$. The reason behind this choice is to assess how the features of the dictionary affect the resulting reconstruction and to identify which combination is the most suitable for the image analysis task (typically, segmentation or detection) one aims to perform next. 
To build the dictionaries, we used a sample of 60 noise-free HR OCT images as a training set and applied the SparseDT approach described in \cite{dictionarySparseDT} with the assumption that the underlying data distribution corresponds exactly to the Cauchy Distribution.
We consider a SR  model with magnification factor $q=4$, so that  $Y\in\R^{256\times 128}$ and $X^{*}\in \R^{1024\times 512}$. 
The FB iterations stop as soon as the tolerance $10^{-5}$ is satisfied or whenever the maximum number of 300 iterations is reached. 
Recalling the condition \eqref{eq:conditions},
 we set $\tau=1.01$, 
 we let $\lambda$ vary in the range $[10^{-7},1]$.
Table \ref{tab::errors} reports the PSNR and SSIM values of the solutions obtained by the possible combinations of the 4 dictionaries and the 3 penalties for two test images which we refer to as OCT1 and OCT2. 
\begin{table}
    \centering
    \resizebox{0.35\textwidth}{!}{%
    \begin{tabular}{|c||a|a||d|d|}
    \hline
    &\multicolumn{2}{|c|}{\textbf{OCT1}}&\multicolumn{2}{|c|}{\textbf{OCT2}}\\
    \hline \hline
     & 	  \textbf{PSNR} &  \textbf{SSIM} & 	  \textbf{PSNR} &  \textbf{SSIM} \\
     \hline\hline
     &\multicolumn{4}{|c|}{Cauchy}\\
     \hline\hline
	$\Dm(256,300)$   	 	& 19.923 			& 0.220			& 19.560 			& 0.200				\\
    $\Dm(64,300)$    	 	& 20.079 			& 0.302			& 19.893 			& 0.224				\\
    $\Dm(256,600)$ 	 		& 19.059 			& 0.188			& 18.694 			& 0.167				\\
    $\Dm(64,600)$   		& \textbf{20.395}	& 0.339			& \textbf{20.200} 	& 0.322				\\
     \hline\hline           
     & \multicolumn{4}{|c|}{MCP}\\
     \hline\hline           
    $\Dm(256,300)$  		& 19.871 			& 0.205			& 19.512 			& 0.186				\\
    $\Dm(64,300)$  	 		& 20.124 			& 0.218			& 19.747 			& 0.197 				\\
    $\Dm(256,600)$  		& 20.386 			& 0.341			& 17.657 			& 0.203 				\\
    $\Dm(64,600)$   		& 20.386 			& 0.341			& 20.148 			& 0.257				\\
     \hline\hline           
     &\multicolumn{4}{|c|}{$\ell_1$}\\
     \hline\hline           
    $\Dm(256,300)$			& 19.928 			& 0.221			& 19.564 			& 0.202				\\
    $\Dm(64,300)$  			& 20.288 			& 0.250 			& 20.001 			&  0.238 			\\
    $\Dm(256,600)$ 			& 19.259 			& 0.163			& 19.300 			& 0.146				\\
    $\Dm(64,600)$  			& 20.376 			& \textbf{0.342}& 20.185 			& \textbf{0.325}			\\

     \hline
    \end{tabular}}
    \caption{PSNR and SSIM values for the two test images.}
    \label{tab::errors}
\end{table}
We notice that the best quantitative results are obtained in correspondence of the dictionaries with $n_p=64$, while, on the contrary, the dictionaries with $n_p = 256$ generally lead to lower quantitative values, but more visually satisfying results, especially in view of further processing of the image such as segmentation and detection.  
We report the results in Fig. \ref{fig:TX12:rec} where we frame some details in yellow, blue and green show respectively some cells suspended in the vitreous, the separation between the vitreous and the retina and finally a portion of the upper layer region.
We remark that the combination of $\Dm(256,600)$ and a non-convex penalty achieves remarkable results in separating the layers of the retinal structure, with Cauchy being the one  better removing background noise at the cost of a slight smoothing in the layer region, which can help layer classification analysis. We support our discussion reporting in Fig. \ref{fig:test1:segm} the $k$-means segmentation with $k=3$ of the reconstructed images for OCT2. The use of this approach for the accurate analysis of layers is left for future research.

\section{Conclusions}
We considered a non-convex variational model for patch-based SR of highly-degraded real murine OCT images promoting sparsity w.r.t. to a heavy-tailed Cauchy dictionary. We enforce sparsity by means of separable non-convex parametric regularisations and assess the performance of the SR model both quantitatively and visually. By imposing a condition on the model/algorithmic parameters preserving the overall non-convexity of the composite functional, 
we guarantee the existence of a unique proximal point. Possible extensions involve generalisation to other regularisers and application to the analysis of OCT data.

\small{\section{Compliance with Ethical Standards}
All procedures were conducted under licence from the UK Home Office and approved by the University of Bristol Ethical Review Group. The study complied with the Association for Research in Vision and Ophthalmology Statement for the Use of Animals in Ophthalmic and Visual Research.
 }


\bibliographystyle{IEEEbib}
\bibliography{refs}

\end{document}